\addtocontents{}{}
\documentclass[11pt]{article}
\usepackage{amsmath,amssymb,color,graphics,epsfig,cite,footnote}
\usepackage{subfigure}
\usepackage{graphicx}
\usepackage{amsfonts}

\newcommand{\hoch}[1]{$\, ^{#1}$}

\newcommand{\be}{\begin{equation}}
\newcommand{\ee}{\end{equation}}
\newcommand{\bea}{\setlength\arraycolsep{2pt} \begin{eqnarray}}
\newcommand{\eea}{\end{eqnarray}}
\newcommand{\nn}{\nonumber}

\def\ft#1#2{{\textstyle{\frac{\scriptstyle #1}{\scriptstyle #2} } }}
\def\fft#1#2{{\frac{#1}{#2}}}

\def\0{{\sst{(0)}}}
\def\1{{\sst{(1)}}}
\def\2{{\sst{(2)}}}
\def\3{{\sst{(3)}}}
\def\4{{\sst{(4)}}}
\def\5{{\sst{(5)}}}
\def\6{{\sst{(6)}}}
\def\7{{\sst{(7)}}}
\def\8{{\sst{(8)}}}
\def\sst#1{{\scriptscriptstyle #1}}

\begin{document}

	
\begin{center}
		
{\Large {\bf Echoes from Classical Black Holes}}
		
\vspace{20pt}
		
{\large Hyat Huang\hoch{1}, Min-Yan Ou\hoch{1}, Meng-Yun Lai\hoch{1}, H. L\"u\hoch{2}}

\vspace{10pt}

{\it \hoch{1}College of Physics and Communication Electronics, \\
	Jiangxi Normal University, Nanchang 330022, China}

\bigskip

{\it \hoch{2}Center for Joint Quantum Studies and Department of Physics,\\
School of Science, Tianjin University, Tianjin 300350, China}

\vspace{40pt}

\underline{ABSTRACT}
\end{center}

We study echoes of free massless scalar from dyonic black holes in Einstein-Maxwell gravity extended with the quasi-topological electromagnetic term. These asymptotically-flat black holes all satisfy the dominant energy condition. We find that for suitable parameters, the effective potential can have double peaks and consequently echoes of quasinormal modes arise. In general the echo frequency associated with the initial wave packet released inside the two peaks is about twice the frequency of the one released outside.

\vfill {\footnotesize hyat@mail.bnu.edu.cn\ \ \ ouminyann@163.com\ \ \ mengyunlai@jxnu.edu.cn\ \ \ mrhonglu@gmail.com}

\newpage
	


\section{Introduction}\label{se1}

The discovery of gravitational waves (GWs) has launched an extraordinary new era in black hole physics. For the merging of two massive objects, the signals of late-time ringdowns encrypt important information that is worth studying. After the initial wave burst of the perturbation, the ringdown enters the stage of quasinormal modes (QNMs) that no longer depend on the initial disturbance; instead, they reflect the properties of the spacetime geometry. Therefore it appears that QNMs are the best to study the black hole geometry that is characterized by having an event horizon. However, it was shown in ref.~\cite{Cardoso:2016rao} that QNMs are in fact related to the light rings of the massive objects, which are not necessarily black holes.  It is therefore hard to distinguish black holes from other horizonless massive objects simply by studying their QNMs \cite{Berti:2009kk,Mark:2017dnq}. In fact there are many exotic compact objects (ECOs) that can mimic the black holes, such as wormholes \cite{Huang:2019arj,Huang:2020qmn,Kanti:2011jz,Kleihaus:2014dla}, boson stars \cite{Schunck:2003kk,Kleihaus:2005me,Hartmann:2010pm} and firewalls \cite{Almheiri:2012rt,Kaplan:2018dqx}.

Data analysis suggests that the recent LIGO observation appears to have detected the ``echo'' signals in the binary black hole waveforms \cite{Abedi:2016hgu,Abedi:2017isz}. However, the study of QNMs in classical black holes indicates that it is hard to create such signals. It was suggested in \cite{Cardoso:2016rao,Cardoso:2017cqb,Cardoso:2016oxy} that echo signals could be used to distinguish the source of gravitational waves: whether it is from a black hole or not. Indeed recent researches in the perturbations of massive objects showed that echo signals could easily arise in phantom wormholes \cite{Bueno:2017hyj,Bronnikov:2019sbx,Liu:2020qia, Ou:2021efv}. In fact echoes can also come from other compact objects \cite{Konoplya:2018yrp}, and a diverse sources including quark stars \cite{Kartini:2020ffp}, braneworld black holes \cite{Dey:2020lhq}, even from a singularity \cite{Chowdhury:2020rfj}. However, it is hard to create echoes from the known traditional classical black hole geometries in Einstein gravity.

Whether black holes can have echoes thus become an important subject to study. Many ideals were put forward to create black hole echoes. Notable examples include quantum black holes \cite{Saraswat:2019npa,Oshita:2020dox,Wang:2019rcf,Manikandan:2021lko}, black hole with discontinuous effective potential \cite{Liu:2021aqh}, or with local Lorentz symmetry violations \cite{DAmico:2019dnn}. However, there is hitherto no example of classical black holes that can produce echoes. By ``classical black holes'', we refer to those constructed in Einstein gravity with minimally-coupled matter. Of course, one can always write black hole metrics with desired properties. The difficulty is that we should require they satisfy some proper energy conditions, such as the dominant energy condition (DEC), or at least the weak energy condition (WEC).

In this paper, we draw the inspiration from the observation of the relation between the QNMs and the light rings, or photon spheres for spherically symmetric configurations. In order to produce echoes, the effective potential should have double peaks so that signals can resonate between the peaks. This feature appears to be closely related to black holes with multiple photon spheres. Such black holes indeed exist in Einstein-Maxwell gravity extended with quasi-topological terms constructed from the Maxwell field strength \cite{Liu:2019rib,Cisterna:2020rkc}. The black holes all satisfy DEC, but can violate the strong energy condition (SEC). The violation of SEC relaxes the condition that gravity must be attractive, and consequently a stable photon sphere can exist, sandwiched between two unstable ones \cite{Liu:2019rib}. We find, as expected, that echoes can indeed emerge from these classical black holes.

In this paper, we study the free massless scalar perturbation, since we would like to use it to probe the spacetime geometry. The scalar does not interact directly with the Maxwell field, which for our purpose here merely provides the proper energy momentum tensor to support the black hole. As was suggested in \cite{Liu:2021aqh,Liu:2020qia}, we can study such scalar perturbation as a proxy for gravitational perturbation. The echoes of scalar and electromagnetic perturbations from wormholes and braneworld black holes were investigated in refs.\cite{Liu:2020qia,Dey:2020pth}.

The paper is organized as follows.  In section 2, we give a general setup for the massless scalar perturbation and the numerical procedure.  In section 3, we review the black holes constructed from the quasi-topological electromagnetism and analyse the effective potentials of the scalar wave equation. In section 4, we give our main results of QNMs that describe black hole echoes.  We conclude the paper in section 5.  In the appendix, we compile various plots of our numerical results.

\section{The general setup}

\subsection{Massless scalar and the effective potential}

In this paper, we consider only the static and spherically symmetric black holes, whose most general metric takes the form,
\be\label{metric}
ds^2=-h(r)dt^2+\fft{dr^2}{f(r)}+r^2d\Omega^2_2\,.
\ee
We assume that the outer horizon of the black hole is located at $r_+$, with $h(r_+)=f(r_+)=0$. To study the black hole echoes due to the metric only, we consider the free massless scalar perturbation, whose equation is
\be\label{formsclar}
\Box \psi\equiv \ft{1}{\sqrt{-g}}\partial_\mu(\sqrt{-g}g^{\mu\nu}\partial_\nu\psi(t,r,\theta,\phi))=0.
\ee
For the background metric \eqref{metric}, the equation can be solved straightforwardly using the separation of variables:
\be
\psi(t,t,\theta,\phi)=\Sigma_{l.m}\ft{\Phi(t,r)}{r} Y_{l,m}(\theta,\phi),
\ee
where $Y_{l,m}$ is the spherical harmonics. Under this, the scalar equation \eqref{formsclar} reduces to
\be\label{eom1}
 -\frac{\partial^2 \Phi(t,r)}{\partial t^2} + {hf} \frac{\partial^2 \Phi(t,r)}{\partial r^2} +\ft{1}{2}( f h^{\prime}+h f^{\prime}) \frac{\partial \Phi(t,r)}{\partial r}
 -\bigg(\ft{l(l+1)}{r^2}h+\ft{1}{2r}(hf)'\bigg) \Phi(t,r)=0,
\ee
where a prime denotes a derivative with respect to $r$. The radial coordinate $r$ lies in the region $(r_+, +\infty)$. For QNMs, ingoing and outgoing boundary conditions should be respectively imposed on the horizon $r_+$ and at the asymptotically infinity. Therefore QNMs do not probe the spacetime geometry inside the event horizon. It is instructive to use the tortoise coordinate to map the the radial region to $(-\infty, \infty)$. Specifically,  we define the tortoise radial coordinate $r_*$ by
\be
dr_*=\fft{dr}{\sqrt{hf}}\,.
\ee
The equation \eqref{eom1} can be cast as
\be\label{eqsolve}
-\fft{\partial^2\Phi}{\partial t^2}+\fft{\partial^2\Phi}{\partial r^2_*}-V(r)\Phi=0,
\ee
where the effective potential $V$ is given by
\be\label{effV}
V(r)=\fft{l(l+1)}{r^2}h+\fft{1}{2r}(hf)'.
\ee
It is clear that after imposing the boundary condition, the property of the function $\Phi$ is solely determined by the effective potential $V$. It is worth noting that the tortoise coordinate transformation can stretch the potential but it does not change its shape, characterized by the number of extrema.

It can be easily seen that we have $V(r_+)=0$ and $V'(r_+)>0$. Furthermore at asymptotically infinity, we have $V(\infty)\rightarrow 0^+$ and $V'(\infty)\rightarrow 0^-$, assuming that $h\sim f\sim 1-2M/r + \cdots$, where $M$ is the mass of the black hole. It follows that the potential $V(r)$ must have at least one maximum. In principle there can exist multiple peaks, with an odd number of total extrema.  Typical black holes, including the Schwarzschild and the Reissner-Nordstr\o m (RN) black holes, $V(r)$ has one extremum, which is the maximum, in which case, we do not expect to have any echoes, since there is no mechanism for resonance.  In order to have echoes, we must have at least three extrema, with a minimum sandwiched between the two maxima.  Such configurations of the effective potential arise naturally for wormholes, but are hard to come by in the case of black holes.  It is thus natural to explore the conditions for which the potential \eqref{effV} has at least double peaks.

One clue is provided by the observation that for large $l$, the leading term of the effective potential is dominant, i.e. $V\sim h/r^2$, which is precisely the effective potential for determining the photon spheres. The extrema of $h/r^2$ with positive $h$ gives the radii of the photon spheres, in the static and spherically symmetric spacetimes.  The existence of multiple peaks of our effective potential is then closely related to the existence of multiple photon spheres.  Classical black holes with multiple photon spheres are rare, but they do exist and we shall discuss this in section \ref{sec:quasi}.

\subsection{Numerical method}

In this paper, we would like to solve wave equation \eqref{eqsolve} with the effective potential \eqref{effV} and obtain the time domain profile of $\Phi(r,t)$. We adopt the the finite difference method, which is effective dealing with the QNMs of both black holes\cite{Zhu:2014sya} and wormholes \cite{Liu:2020qia}.

To obtain the finite difference equation, we discretize the coordinates $t=i \Delta t$ and $r_*=j \Delta r_*$, where the $i$ and $j$ are integers. The scalar field $\Phi$ and the effective potential $V$ are also discretized:
\be
\Phi(t,r)=\Phi_{i,j}\equiv(i\Delta t,j\Delta r_*)\,,\qquad
V(r_*)=V_j\equiv V(j \Delta r_*)\,.
\ee
With this, the differential equation \eqref{eqsolve} becomes a set of iterative algebraic equations:
\bea\label{dis1}
&&-\frac{\Phi(i+1,j)-2\Phi(i,j)+\Phi(i-1,j)}{\Delta t^2}
+\frac{\Phi(i,j+1)-2\Phi(i,j)+\Phi(i,j-1)}{\Delta r_*^2}\\
&&\qquad\qquad-V(j)\Phi(i,j)=0\,.\nn
\eea
Our goal is to solve the above equations, starting from a certain initial wave packet. For convenience, we set it in Gaussian distribution, i.e.
\be
\psi(t=0,r_*)=e^{- \frac{(r_*- \bar{a})^2}{2}}\,,\qquad \hbox{and}\qquad  \psi(t<0,r_*)=0\,.
\ee
The parameter $\bar{a}$ is the location of the center of the initial wave packet. Two inequivalent situations can be considered.  One is that the wave packet is outside the double peaks. In this case we find the echoing QNMs have the same characteristics whether it is put close to the horizon or out away from the peaks.  We therefore present the results only for the latter case.  Another intriguing situation is when we put the initial wave packet in the potential well between the two peaks.  We find when the releasing location $\bar a$ is closer to the bottom of the potential well, the echo signals will become more distinguishable. Furthermore, the echo frequency, or the number of echoes, roughly doubles the one associated with the wave packet put outside the peaks.

The boundary conditions and \eqref{dis1} are now given by
\bea\label{dis2}
&&\Phi(t,r_*)|_{r_*\to -\infty}=e^{-{\rm i}{\omega r_*}},\qquad\Phi(t,r_*)|_{r_*\to +\infty}=e^{{\rm i}{\omega r_*}},\nn\\
&&\Phi(i+1,j)=-\Phi(i-1,j)\nn\\
&&+\bigg(2-2 \frac{\Delta t^2}{\Delta r_* ^2}- \Delta t^2 V(j) \bigg) \Phi(i,j)
+ \frac{\Delta t^2}{\Delta r_* ^2}  \bigg(\Phi(i,j+1)+\Phi(i,j-1) \bigg),
\eea
where $\omega$ is the frequency of the QNMs. (We should distinguish the frequency of the QNMs from the echo frequency.)  In fact, it is a free parameter in our numerical calculations. The reason is that the signal has not reached the bound of the grids. According to the von Neumann stability condition, we set $\Delta t / \Delta r_*=0.5$ in our numerical calculations. More details are referred to ref.~\cite{Zhu:2014sya}.

\section{Black holes and double-peak potentials}
\label{sec:quasi}

\subsection{Black holes from quasi-topological electromagnetism}

In this section, we review Einstein-Maxwell gravity extended with a quartic-order quasi-topological term constructed from the Maxwell field strength $F=dA$. The Lagrangian is \cite{Liu:2019rib}
\be
{\cal L}=\sqrt{-g}\Big(R- F^{\mu\nu} F_{\mu\nu}-\alpha\,\big((F^{\mu\nu} F_{\mu\nu})^2-2F^\mu_\nu F^\nu_\rho F^\rho_\sigma F^\sigma_\mu\big)\Big)\,,\label{d4lag}
\ee
where $\alpha$ is the coupling constant of the quasi-topological term. The theory admits an exact solution of asymptotically-flat dyonic black holes that are both static and spherically symmetric:
\be\label{sol}
ds^2 = - h dt^2 + \fft{dr^2}{h} + r^2 \big(d\theta^2 + \sin^2\theta\, d\varphi^2\big)\,,\quad
F= - \fft{q r^2}{r^4 + 4 \alpha p^2}dt\wedge dr + p \sin\theta d\theta\wedge d\varphi\,,
\ee
with
\be
 h= 1 -\frac{2M}{r}+\frac{p^2}{r^2}+\frac{q^2}{r^2} \, _2F_1\Big(\frac{1}{4},1;\frac{5}{4};-\frac{4 \alpha\, p^2}{r^4}\Big)\,.\label{fsol}
\ee
The general solution contains three integration constants, the mass $M$, electric and magnetic charges $(q,p)$. When $\alpha q p=0$, the solution reduces to the usual RN black hole. The property of these black holes were analysed in \cite{Liu:2019rib}. They all satisfy DEC, but can violate SEC.  For suitable parameters, black holes with four horizon can emerge. In this paper, we are interested in black holes with two horizons, but with an unusual feature that the profile function $h$ is not monotonous outside the horizon.  It has an extra wiggle such that the gravity force can vanish and even be repulsive in certain finite region outside the horizon. This feature is a direct consequence of the SEC violation.

\begin{figure}[htbp]
\centering
\subfigure[$M=6.80$]{
\includegraphics[width=0.45\textwidth]{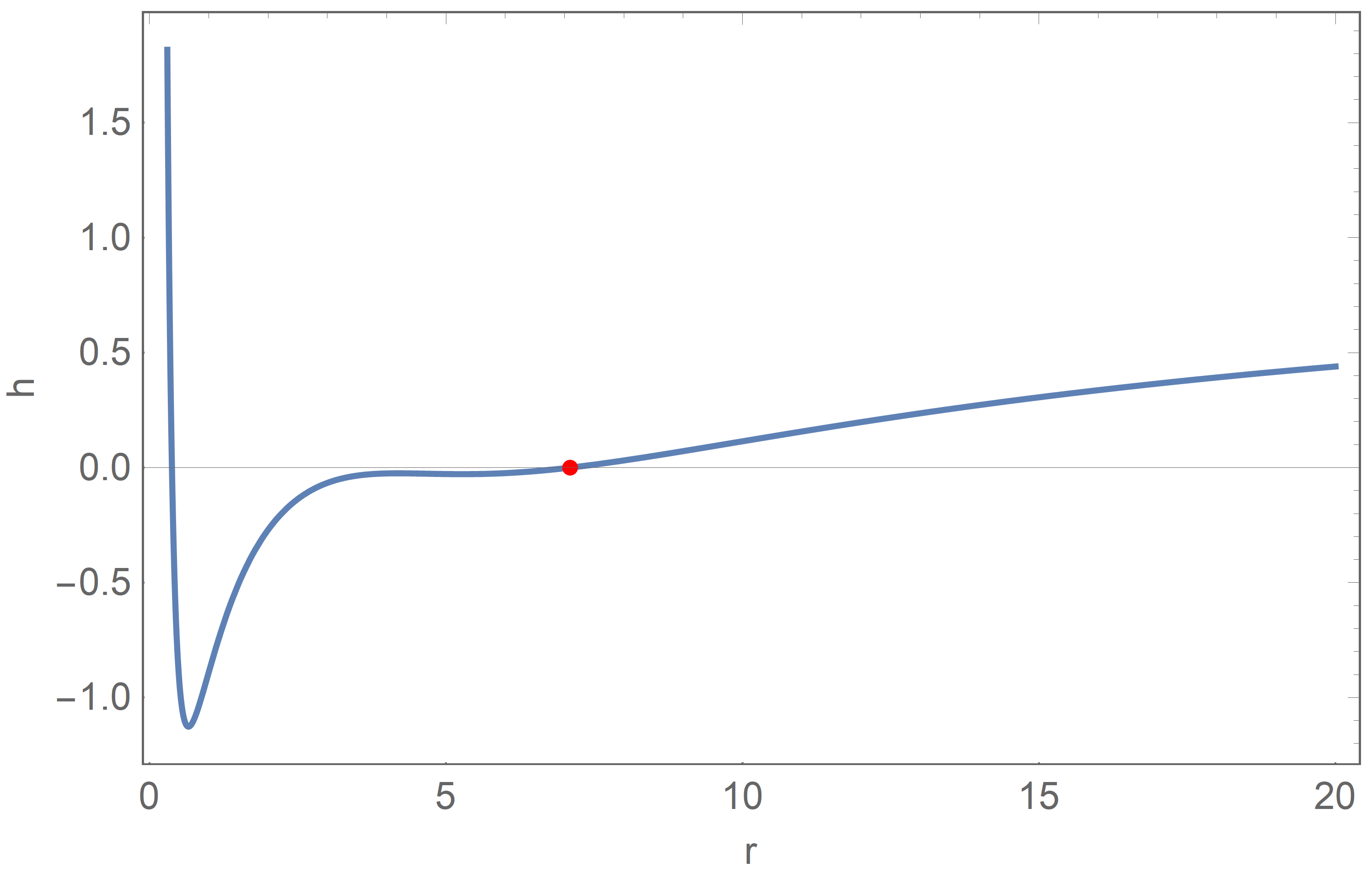}
}
\quad
\subfigure[$M=6.65$]{
\includegraphics[width=0.45\textwidth]{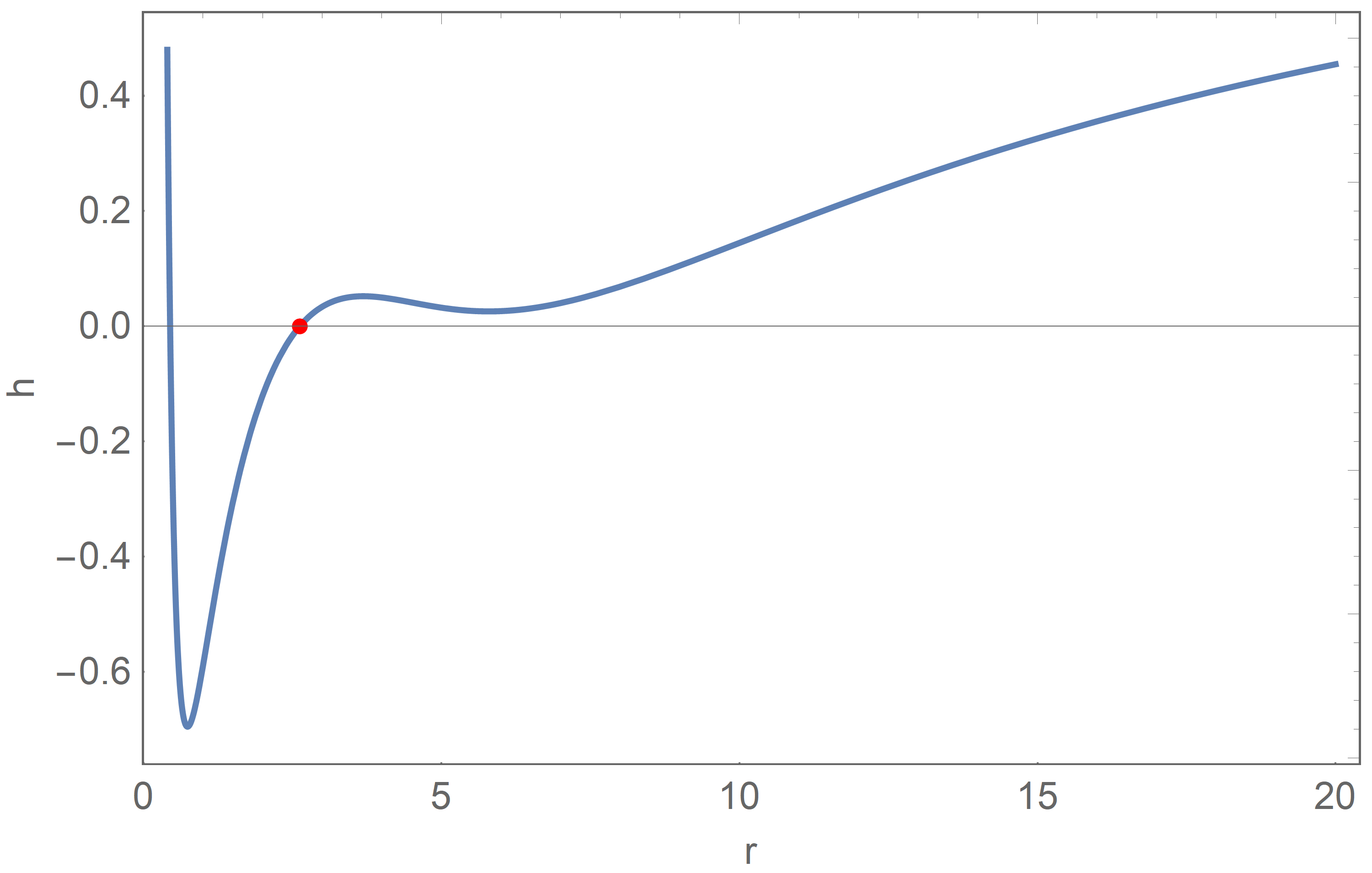}
}
\caption{\it\small  Here we give the metric profile function $h(r)$ for two different masses, with $(\alpha, p, q)=(149.90, 0.945, 6.863)$. Both black holes have two horizons, but depending on the mass, we see that the function $h$ may not be monotonous outside the outer horizon, indicated by the red dot.}\label{hplots}
\end{figure}

For a concrete example, we choose the parameters, e.g. $(\alpha, p, q)=(149.90, 0.945, 6.863)$.
For $M<6.370$, solution is over charged and the curvature singularity at $r=0$ is naked. When $6.355 <M<6.729$, the black hole has two horizons and $h$ is not monotonous outside the outer horizon, but has a wiggle with a local maximum and a minimum. Further increasing the mass such that $6.723<M<6.760$, the local minimum dips down to be negative so that the black hole has four horizons.  When $M>6.760$, the two inner horizons disappear and the black hole reverts back to have two horizons.  In the last two cases, the function $h$ is monotonous outside the outer horizon. In Fig.~\ref{hplots}, we present the plots for the function $h$ for $M=6.80$ and $M=6.65$.

We see from Fig.~\ref{hplots} that both black holes with $M=6.80$ and $M=6.65$ have two horizons.  However, when $M=6.80$, the function $h$ is monotonous outside the horizon, but it has a wiggle when $M=6.65$. This extra wiggle implies that the black hole has three photo spheres, with a stable photon sphere sandwiched between two unstable ones \cite{Liu:2019rib}.  As we have discussed earlier, we expect that such spacetime geometry will give rise the effective potential with double peaks, which we discuss in the next subsection.

\subsection{Effective potentials of black hole solutions}

\begin{figure}[htbp]
\centering
\subfigure[Potential with single peak at $M=6.80$]{
\includegraphics[width=0.45\textwidth]{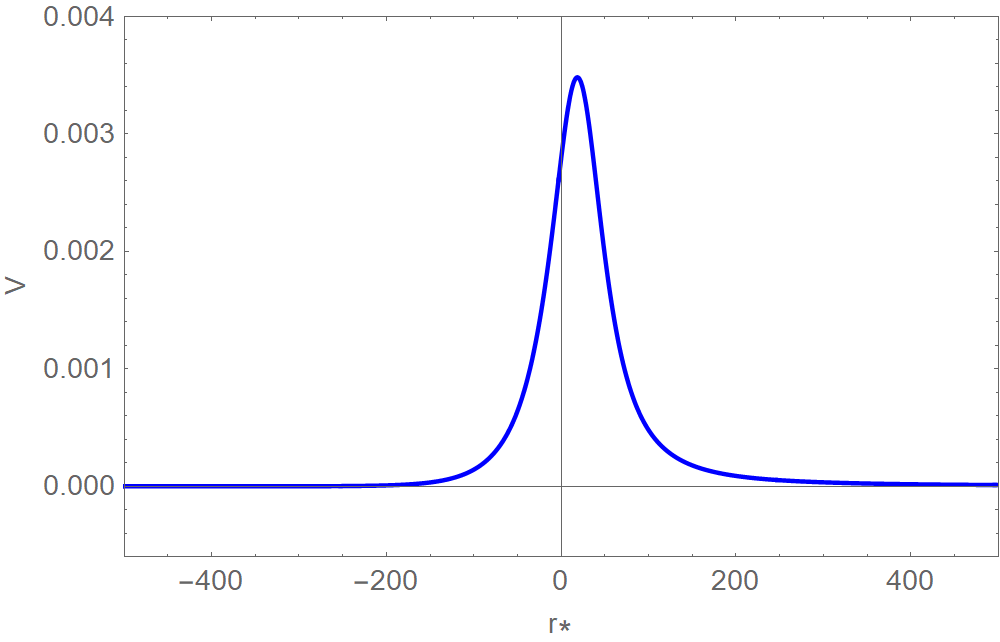}
}
\quad
\subfigure[Potential with double peaks at $M=6.65$]{
\includegraphics[width=0.45\textwidth]{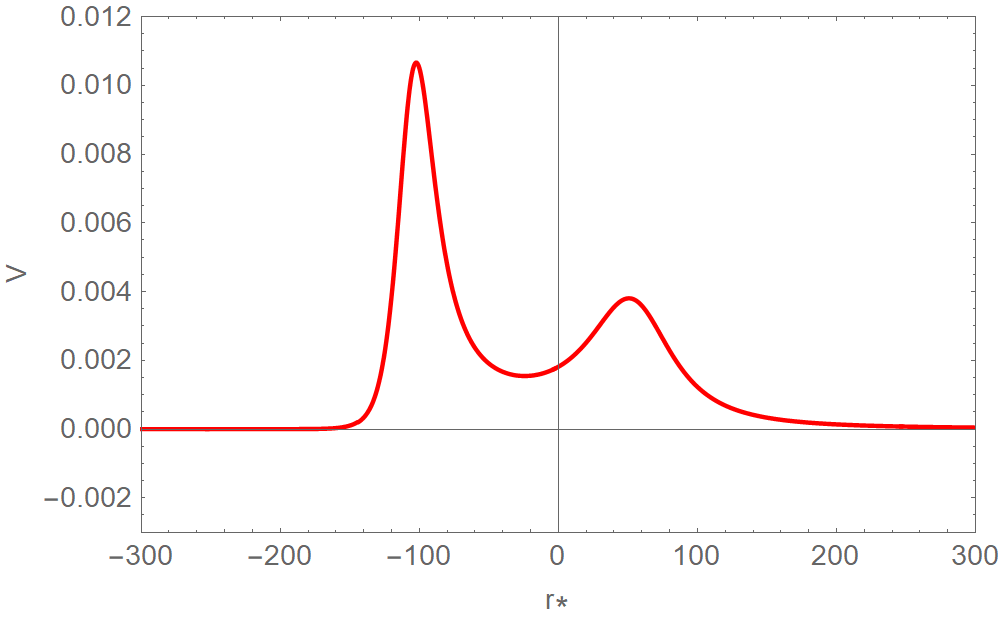}
}
\caption{\it\small  Two typical shapes of the effective potentials are plotted in terms of the tortoise coordinates. The parameters here are chosen to be $\alpha=149.90, p=0.945, q=6.863$ and $l=1$.}\label{shapes}
\end{figure}

The effective potential of the massless scalar wave is given \eqref{effV}. For our case, we have $f=h$, with h given by \eqref{fsol}. We find that there are two types of the effective potentials, depending on the values of the black hole parameters. One involves a single peak, and the other has double peaks. Taking the same values associated with Fig.~\ref{hplots}, we present the corresponding effective potentials using the tortoise coordinates in Fig.~\ref{shapes}.

As we see from Fig.~\ref{shapes} that when $M=6.80$, the effective potential has a single peak, which is typical for many black holes including the Schwarzschild and RN black holes. It has been pointed out in ref.\cite{DAmico:2019dnn} that such black holes can not produce echoes, without introducing a reflective boundary condition by hand.

To produce echoes in classical black holes, it calls for effective potentials with at least two peaks. It is easy to be realized in wormhole  metrics \cite{Bueno:2017hyj,
Bronnikov:2019sbx,Liu:2020qia,Ou:2021efv}. However, there are few such black holes. For the $M=6.65$ black hole, the double-peak potential is achieved by violating SEC such that gravity can be repulsive in certain regions of spacetime. However, since our solutions satisfy DEC, they can nevertheless be models for realistic black holes.

Having established that the double-peak potential can arise, we explore the parameter space around the above example to see how the potential depends on various parameters including the mass, charge and also the $l$ parameter associated with the spherical harmonics.  In order not to interrupt the flow of the main text, we present all the plots of the double-peak potentials in appendix \ref{appa}. The effective potential depends on $(M,q,p,l,\alpha)$. In Fig.~\ref{changingm}, Fig.~\ref{changingq} and Fig.~\ref{changingl}, we give the effective potentials with varying $M$, $q$ and $\ell$ respectively. It should be mentioned that when $l\ne 0$, the effective potential is always nonnegative. However, when $\ell=0$, there may exist regions where the potential becomes negative. We find that in this case the echoes are not discernible. In the next subsection, we study the time-domain profiles of the corresponding QNMs and see that they generally produce echoes.

While the existence of multiple-photon spheres provides the inspiration for double-peak potentials when the first term in \eqref{effV} is dominant, the second term in \eqref{effV} suggests that echoes may happen even when the function $h$ is monotonous, provided that $h'$ is not. We shall present such an example in the next section, together with its echoing QNM.

\section{Numerical echoes from black holes}

In this section, we present our main results, namely the QNMs that carry the echoe signals when the effective potentials have double peaks.  As we mentioned in the setup, there are two situations to consider: (1) the initial wave packet is outside the two peaks; (2) it is between the two peaks.

\subsection{Wave packet outside the double peaks}

\begin{figure}[htbp]
\centering
\subfigure[No echoes from single-peaks potential]{
\includegraphics[width=0.45\textwidth]{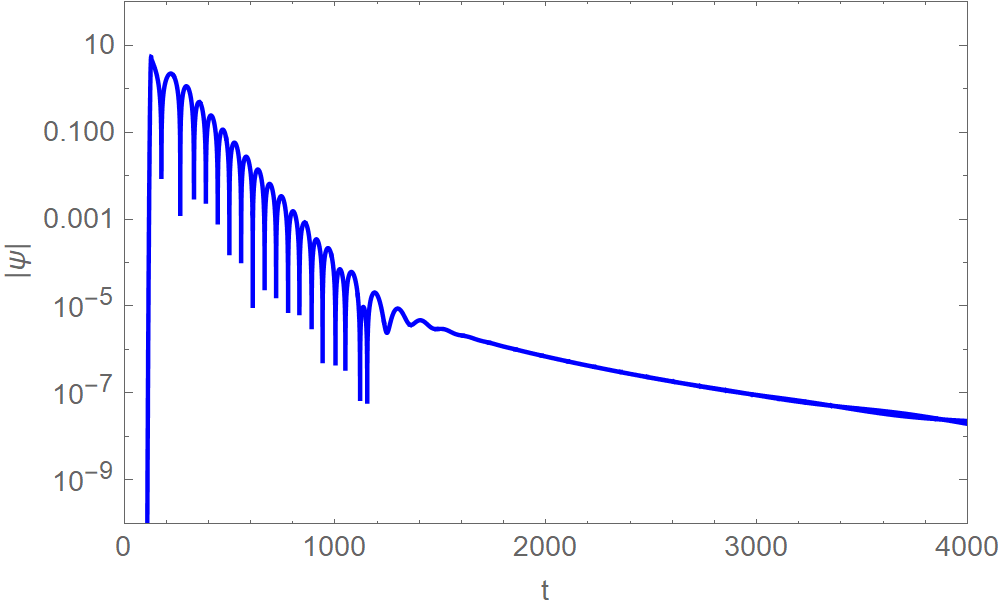}
}
\subfigure[Echoes from double-peaks potential]{
\includegraphics[width=0.45\textwidth]{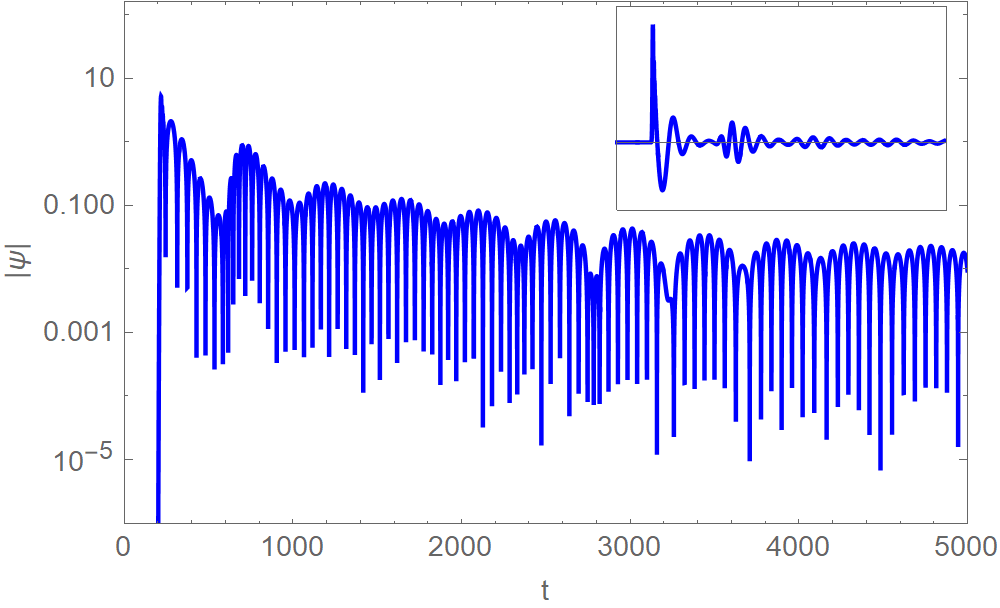}
}
\caption{\it\small These two time-domain profiles correspond to the effective potentials shown in Fig.~\ref{shapes}, where we set $\alpha=149.90, p=0.945, q=6.863, l=1$. The left column has $M=6.8$ while the right column has $M=6.65$. They show that there is no echo for the single-peak potential, while the echoes arise in double-peak potential.}\label{singlepeakQ}
\end{figure}

Here we present the results for the first case. For the purpose of comparison, we first give the QNM when the potential has only one peak. We consider the $M=6.80$ case of Figs.~\ref{hplots}, \ref{shapes}.  As expected, the original wave packet evolves as a damping oscillator, and it produces no echoes, as can be seen by the left column of Fig.~\ref{singlepeakQ}.

For the double-peak potentials, we can observe the echo signals after the initial damping of the wave packet. A concrete example of the time-domain profile of the echoes is presented in right column of Fig.~\ref{singlepeakQ}. It is clear that the echoes arise because of resonant reflection by the two peaks. Therefore the frequency, or the number of echoes does not depend on the releasing position, as long as they are outside the two peaks. (We shall study presently the case when the initial wave packet is released inside the two peaks.)

As we discussed in the previous section, we have obtained many examples of double-peak potentials, by varying parameters such as $M$, $q$ and $l$.  We construct the corresponding QNMs and obtain the time-domain profiles.  We present the results in the appendix \ref{appa} in
Fig.~\ref{Mecho}, Fig.~\ref{qecho} and Fig.~\ref{lecho} respectively.  As we can see, the signature of echo signals depends on many factors and to distinguish them requires precision measurements. One special case is worth mentioning, namely $l=0$. The effective potential can be negative in some regions and the echo signals are hardly discernible. This phenomenon was first noted in the wormhole cases \cite{Churilova:2019cyt}.

From these examples, one might draw a conclusion that the double-peaks potential arises only in the case when $h(r)$ is not monotonically increasing. However, this is not true. We find an example where $h(r)$ is in fact monotonically increasing outside the horizon, but the corresponding effective potential continue to have double peaks. This case arise when we choose the following parameters:
\be
\alpha=123.90\,,\quad p=1.181\,,\quad q=6.863\,,\quad l=3\,,\quad M=6.58\,.
\ee
It is clear that the second term in \eqref{effV} can also give nontrivial contribution since $h'$ is not monotonous is this case.  Our numerical study shows that this black hole can also produce echoes, although the signals are less clear in later time.  We show the results of this example in Fig.~\ref{hdandiao} in the appendix.

\subsection{Wave packet within the double peaks}

\begin{figure}[htbp]
\centering
{
\includegraphics[width=0.5\textwidth]{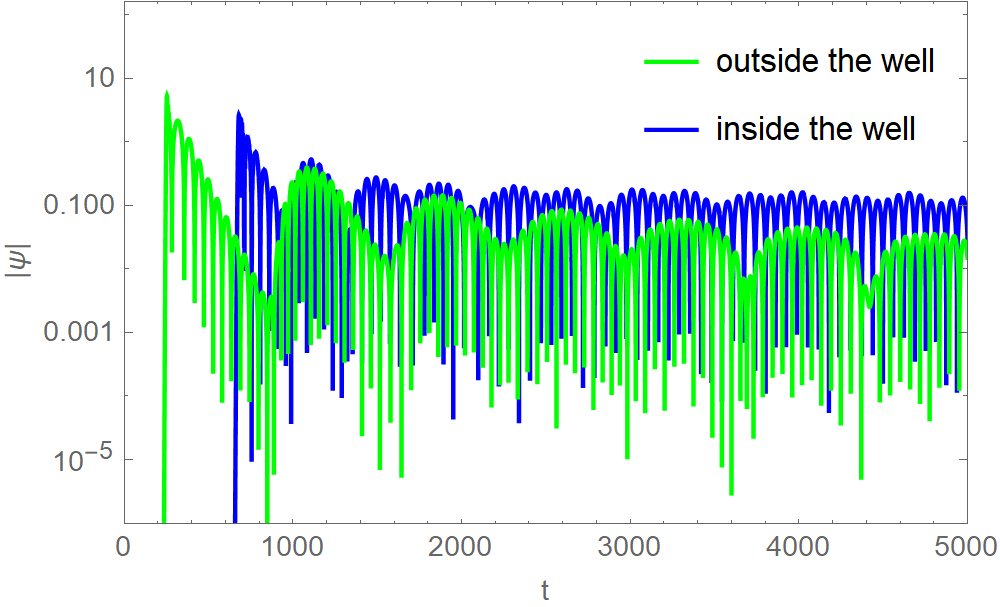}
\caption{\it\small Here are the time-domain profiles of initial wave packets released inside and outside the two potential peaks. The corresponding $V$ is plotted in Fig.~\ref{changingm} with $\alpha=149.90, p=0.945, q=6.86, l=1, M=6.7$. The green lines correspond to the initial wave packet at $r_*=100$ (outside the well). The blue lines correspond to the initial wave packet at $r_*=-76$ (the bottom of the well). The echo frequency of the latter is about twice of the former, and the shapes are less pronounced.}\label{echoinout}}
\end{figure}

In the previous subsection, we focus on echoes of the initial wave packet released outside the two peaks of the effective potential. We now release it in the potential well between the two peaks. As a concrete example, we consider the potential given in Fig.~\ref{changingm}, with
$\alpha=149.90, p=0.945, q=6.863, l=1, M=6.7$. We find that the echo signals become more sensitive to the exact release location. In particular the echoes become more distinguished when the releasing location is at the bottom of the potential well, in which case, we find that the echo frequency nearly doubles. This phenomenon is explicitly depicted in Fig.~\ref{echoinout}, and we see that echoes of the wave packet outside the peak is much more pronounced.

\section{Conclusion}

In this paper, we studied the echoes of a free scalar wave packet initially released outside the dyonic black hole in Einstein-Maxwell gravity extended with quartic quasi-topological terms constructed by the Maxwell field strength. We were motivated by the observation that the phenomenon of black hole echoes was closely related to having multiple photon spheres.  It was well established that these black holes could admit multiple photon spheres, and not surprisingly, our QNM constructions showed that they indeed could have echoes for suitable parameters. Intriguingly we found that when the initial wave packet was released in the bottom of the potential well between the two peaks, its echo frequency is about doubled, compared to the case when the wave packet is released outside the two peaks.  We expect that the similar phenomenon should occur in echoes from non-black hole objects, but it is certainly worth checking.

Our results do not dispute the fact the Schwarzschild or the RN black holes do not have echoes within the framework of general relativity. However, it is important to stress that the black holes studied in this paper all satisfy DEC, and that the free scalar probes only the geometry. In other words, the quasi-topological electromagnetism plays no particular role other than provides the proper matter energy-momentum tensor to curve the spacetime. Our results therefore show the general statement that traditional classical black holes in Einstein gravity coupled to suitable matter satisfying DEC can produce echoes.  An observation of echoes in itself cannot rule out the black hole source, without precision measurements.

The connection between black hole echoes and multiple photon spheres become apparent for large $l$; however, we found that echoes could also arise even when $h$ is monotonous outside the event horizon, when the second term in the effective potential \eqref{effV} becomes nontrivial. Therefore it is far from clear what is the necessary condition for a black hole to echo. It is nevertheless tempting to conjecture that black holes satisfying SEC cannot produce echoes, or more weakly, the no-echo condition requires both SEC and DEC.

\newpage
\section*{Acknowledgement}

We are grateful to Peng Liu, Senping Luo, Bing Sun and Runqiu Yang for useful discussions. H.L.~is supported in part by the National Natural Science Foundation of China (NSFC) grants No.~11875200 and No.~11935009.

\appendix

\section*{Appendix}

\section{Plots of potentials and echoing time-domain profiles}\label{appa}

In this appendix, we compile all the figures mentioned in the main text. For appropriately chosen parameters, we obtain the effective potential $V(r)$ in \eqref{effV} with double peaks and we draw the corresponding echoing QNMs.

\bigskip

\begin{figure}[htbp]
\centering
\subfigure[$V(r)$]{
\includegraphics[width=0.45\textwidth]{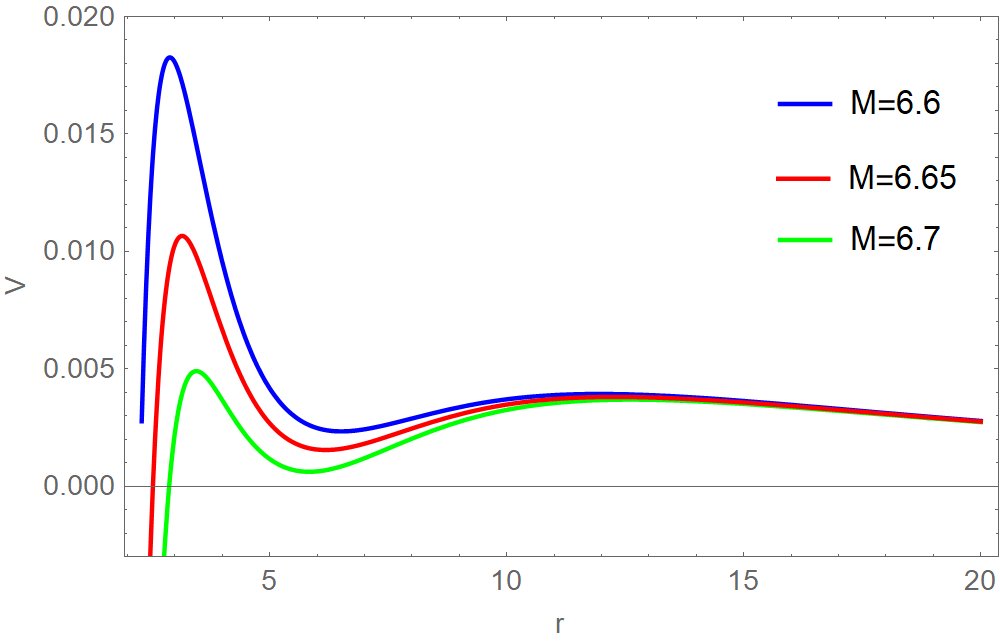}
}
\quad
\subfigure[ $V(r_*)$]{
\includegraphics[width=0.45\textwidth]{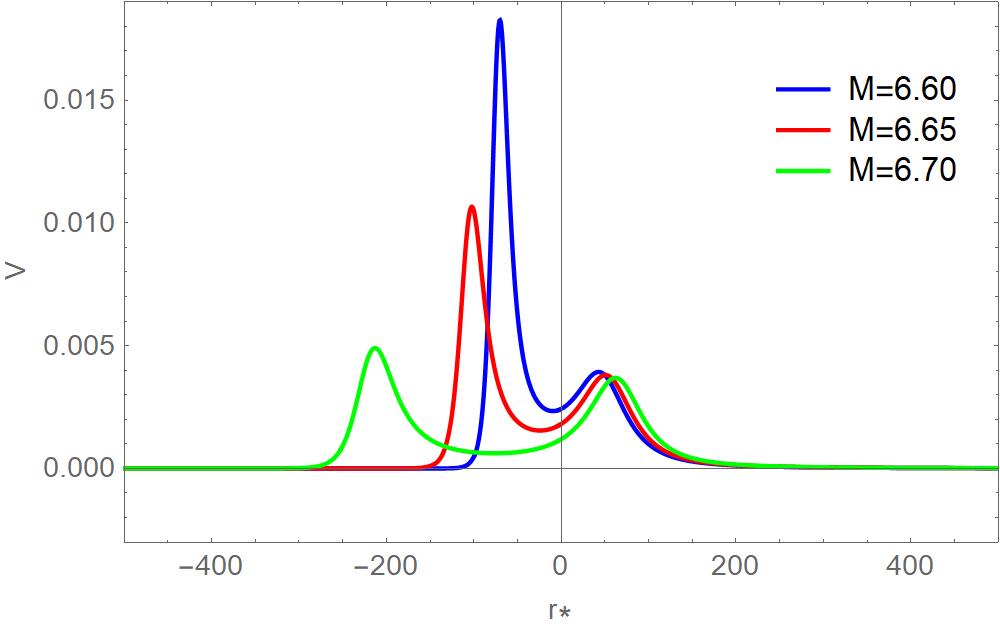}
}
\caption{\it\small Here we fix the parameter $\alpha=149.90, p=0.945, q=6.863, l=1$ and vary the mass $M$. In the left, the potential is plotted as a function of $r$, and the right as a function of $r_*$. The mass lies in a narrow region where $V$ has two peaks. These potentials lead to echo signals presented in Fig.~\ref{Mecho}.}\label{changingm}
\end{figure}

\begin{figure}[htbp]
\centering
\subfigure[$M=6.60$]{
\includegraphics[width=0.45\textwidth]{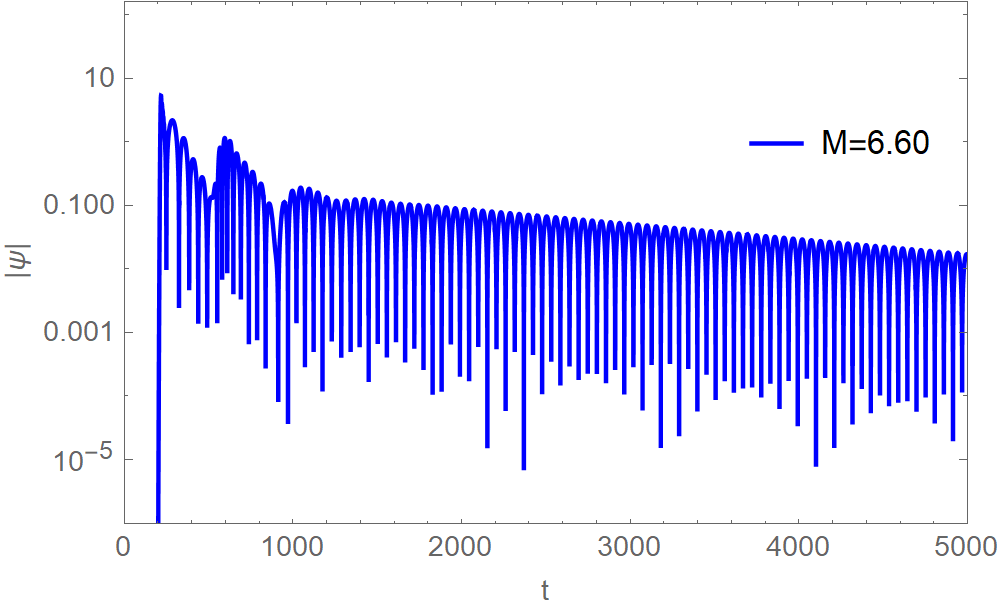}
}
\quad
\subfigure[$M=6.65$]{
\includegraphics[width=0.45\textwidth]{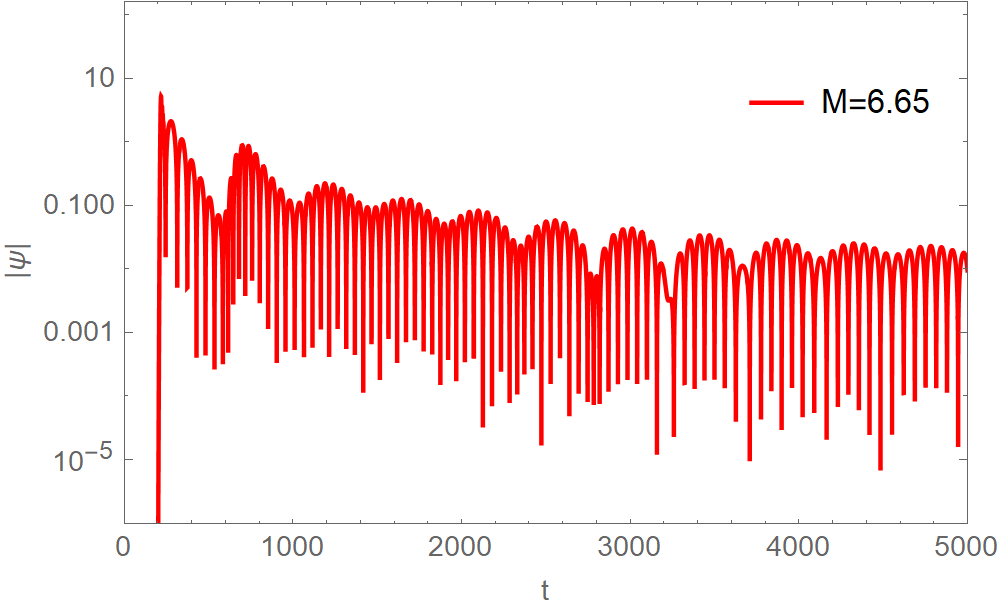}
}
\quad
\subfigure[$M=6.70$]{
\includegraphics[width=0.45\textwidth]{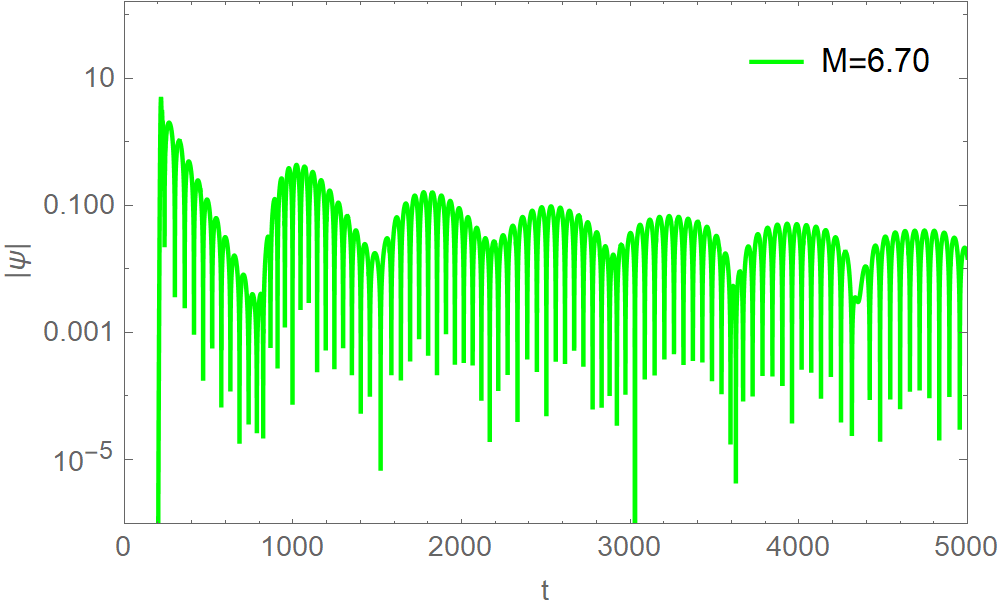}
}
\quad
\subfigure[All together]{
\includegraphics[width=0.45\textwidth]{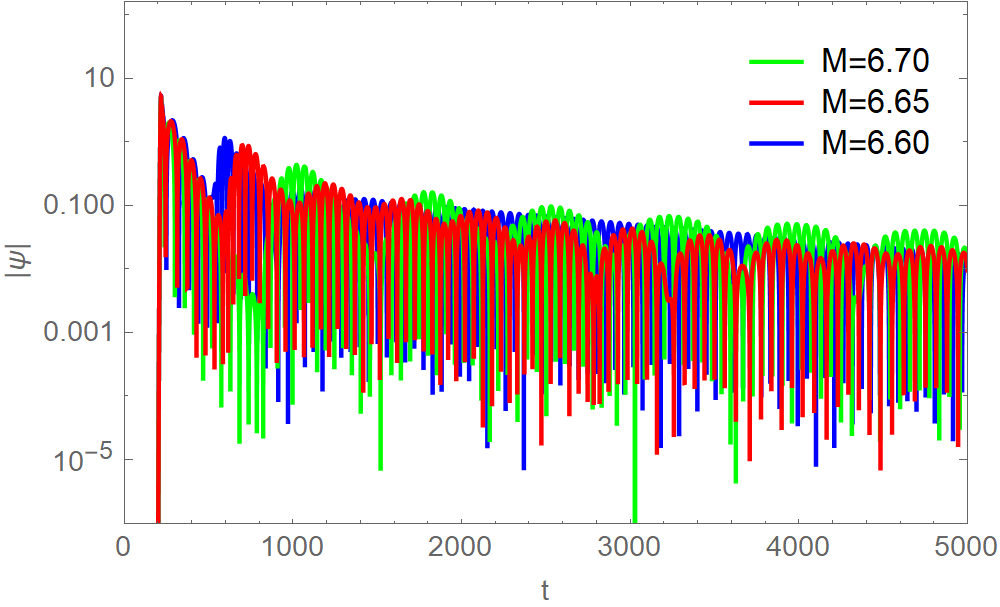}
}
\caption{\it\small These are the time-domain profiles with varying $M$, associated with the effective potentials given in Fig.~\ref{changingm}. The echoes are more pronounced with larger mass in these parameters regions.}\label{Mecho}
\end{figure}

\begin{figure}[htbp]
\centering
\subfigure[$V(r)$]{
\includegraphics[width=0.45\textwidth]{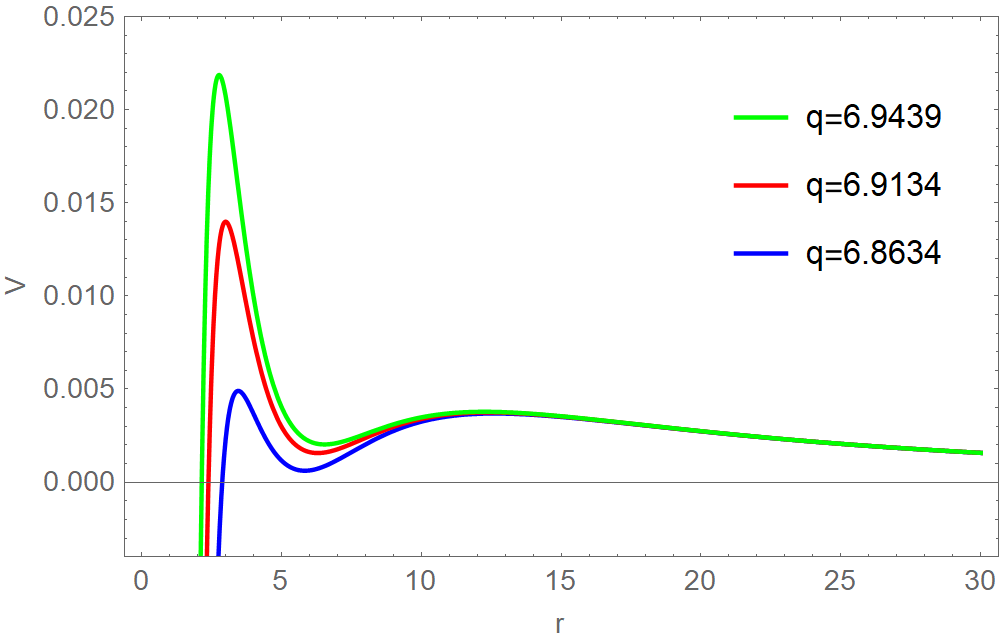}
}
\quad
\subfigure[ $V(r_*)$]{
\includegraphics[width=0.45\textwidth]{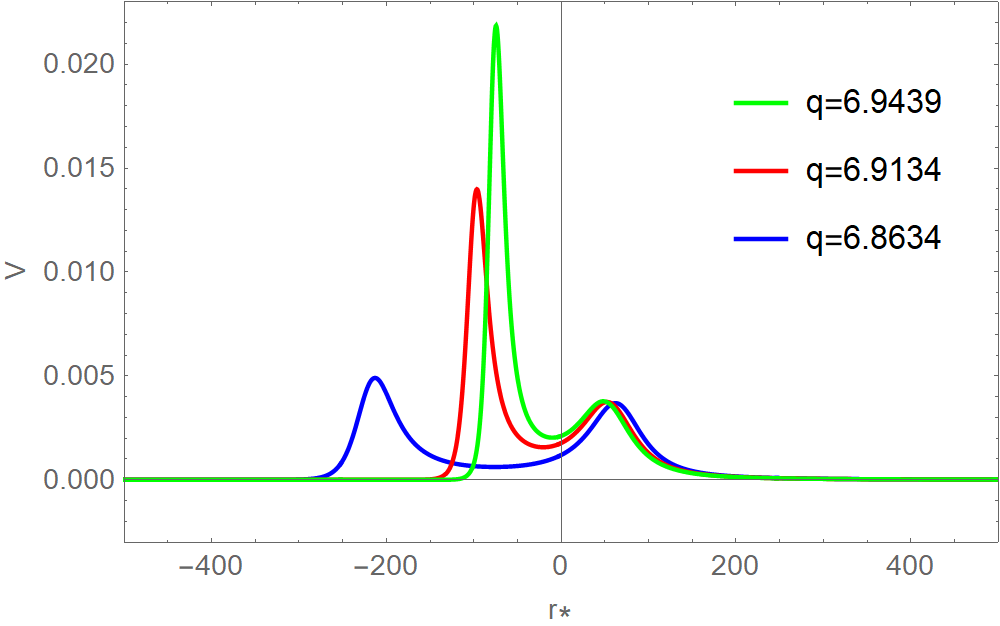}
}
\caption{\it\small In these plots, we fix $M=6.7, \alpha=149.90, p=0.945, l=1$ and study the effective potential by varying the electric charge $q$.}\label{changingq}
\end{figure}

\begin{figure}[htbp]
\centering
\subfigure[$q=6.8634$]{
\includegraphics[width=0.45\textwidth]{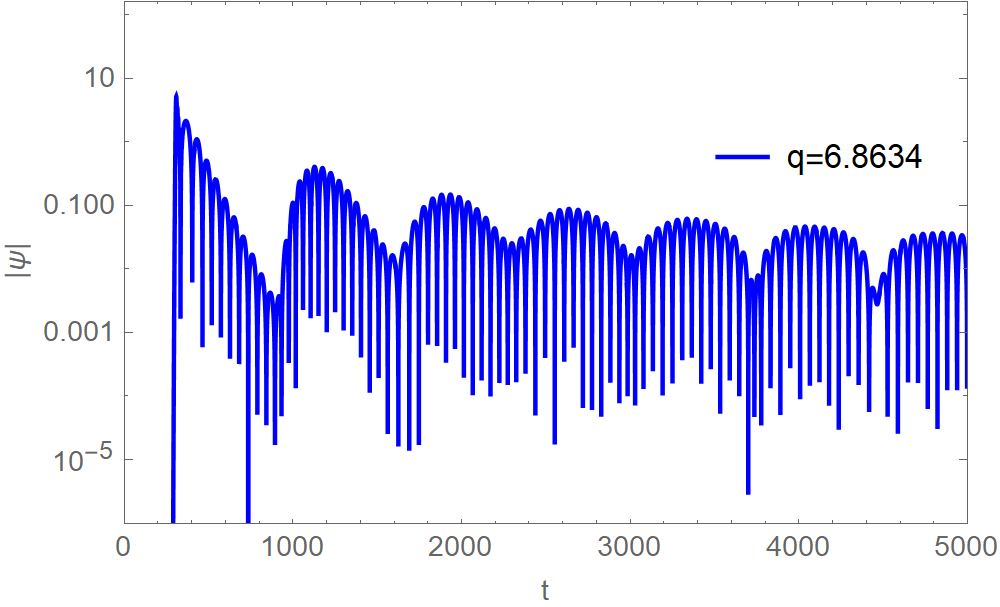}
}
\quad
\subfigure[$q=6.9134$]{
\includegraphics[width=0.45\textwidth]{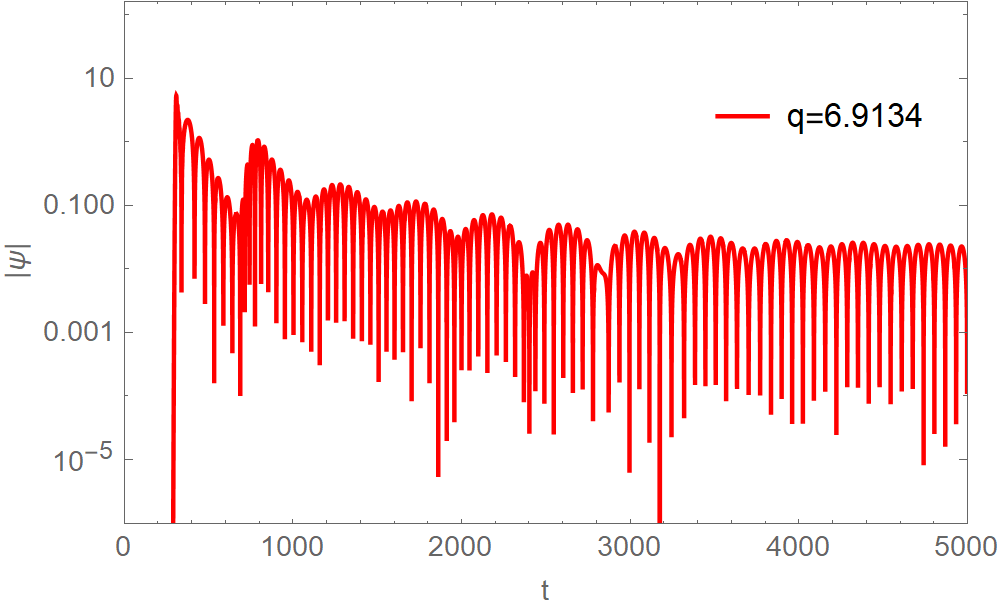}
}
\quad
\subfigure[$q=6.9439$]{
\includegraphics[width=0.45\textwidth]{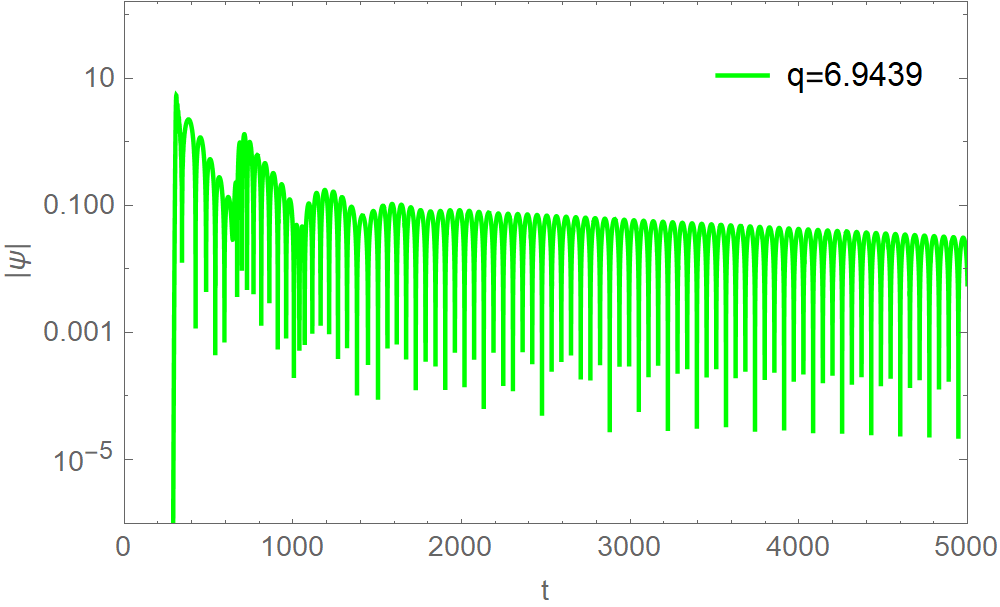}
}
\quad
\subfigure[All together]{
\includegraphics[width=0.45\textwidth]{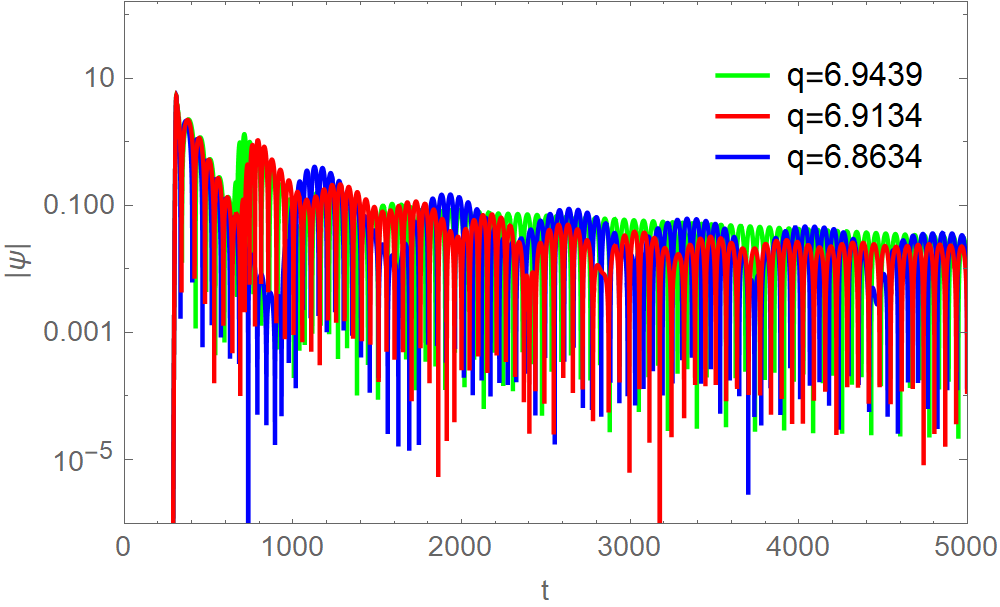}
}
\caption{\it\small These are time-domain profiles with varying $q$, associated with the effective potentials given in  Fig.~\ref{changingq}. The echo is most pronounced for the smallest $q$ in this set of parameters.}\label{qecho}
\end{figure}

\begin{figure}[htbp]
\centering
\subfigure[$l\neq0$]{
\includegraphics[width=0.45\textwidth]{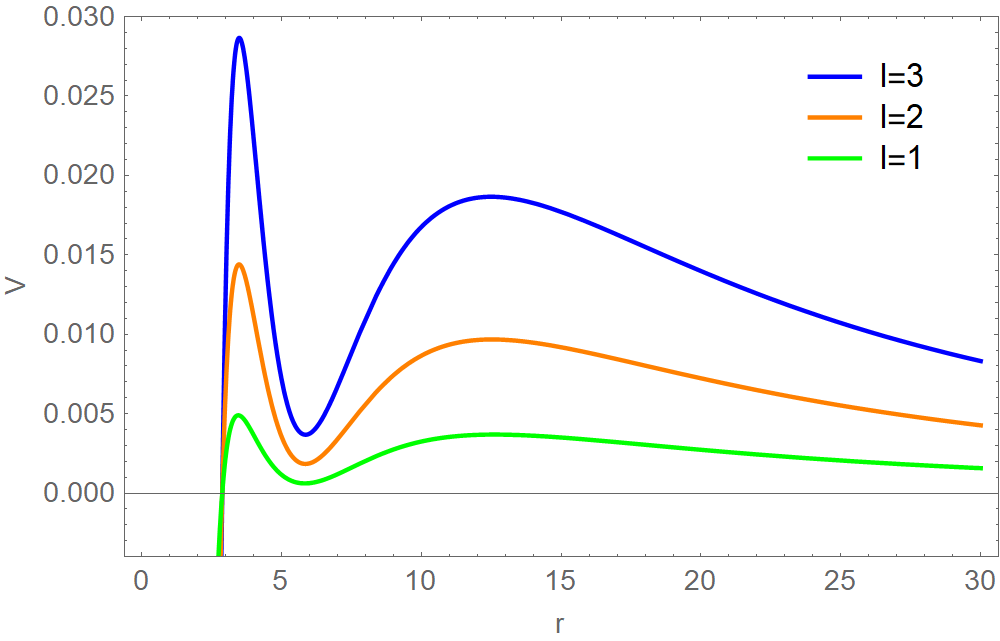}
}
\quad
\subfigure[ $l=0$]{
\includegraphics[width=0.45\textwidth]{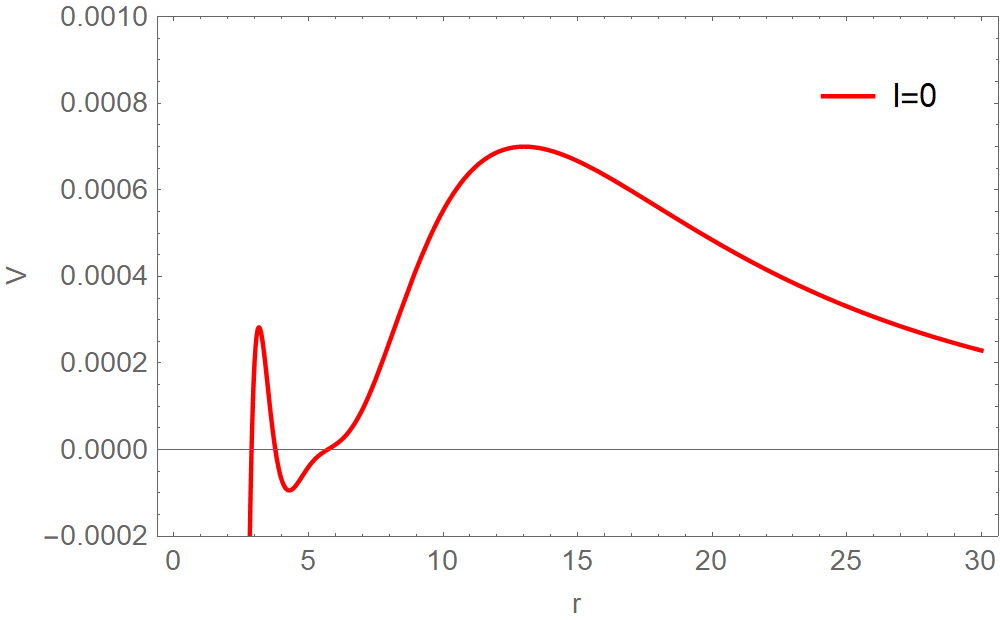}
}
\caption{\it\small  Here we consider varying the parameter $l$, with fixed  $M=6.7, \alpha=149.90, p=0.945, q=6.863$. The left shows $V$ with $l\neq0$; the right with $l=0$. Larger $l$ causes higher peaks. when $l=0$, the potential can be negative in some middle regions. }\label{changingl}
\end{figure}

\begin{figure}[htbp]
\centering
\subfigure[$l=0$]{
\includegraphics[width=0.45\textwidth]{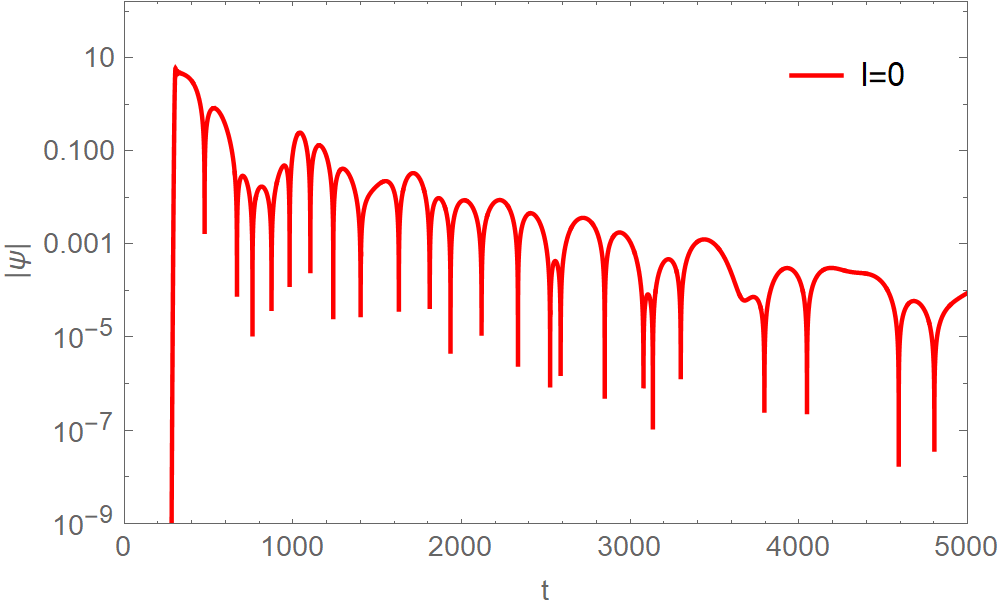}
}
\quad
\subfigure[$l=1$]{
\includegraphics[width=0.45\textwidth]{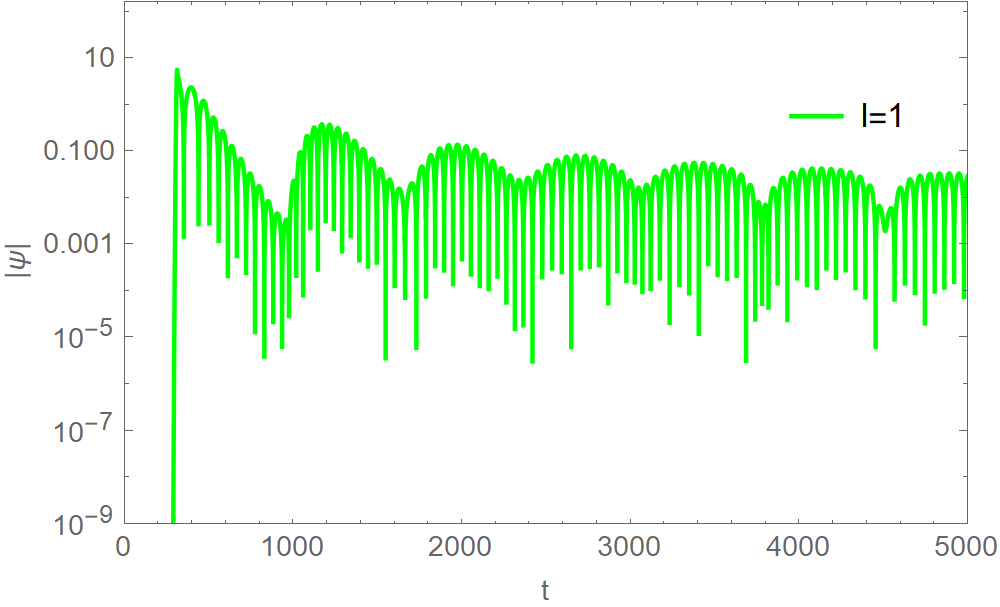}
}
\quad
\subfigure[$l=3$]{
\includegraphics[width=0.45\textwidth]{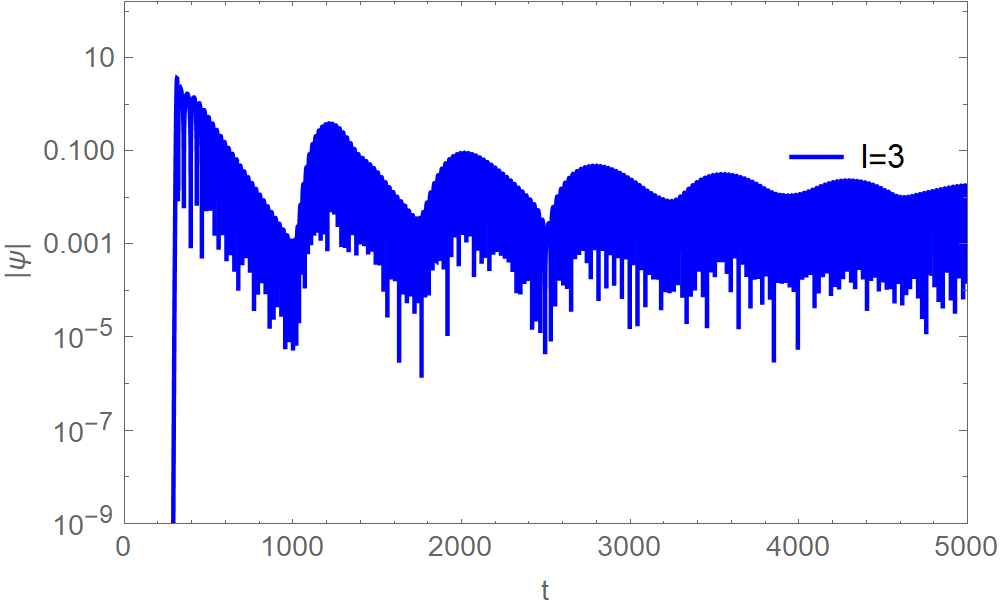}
}
\quad
\subfigure[All together]{
\includegraphics[width=0.45\textwidth]{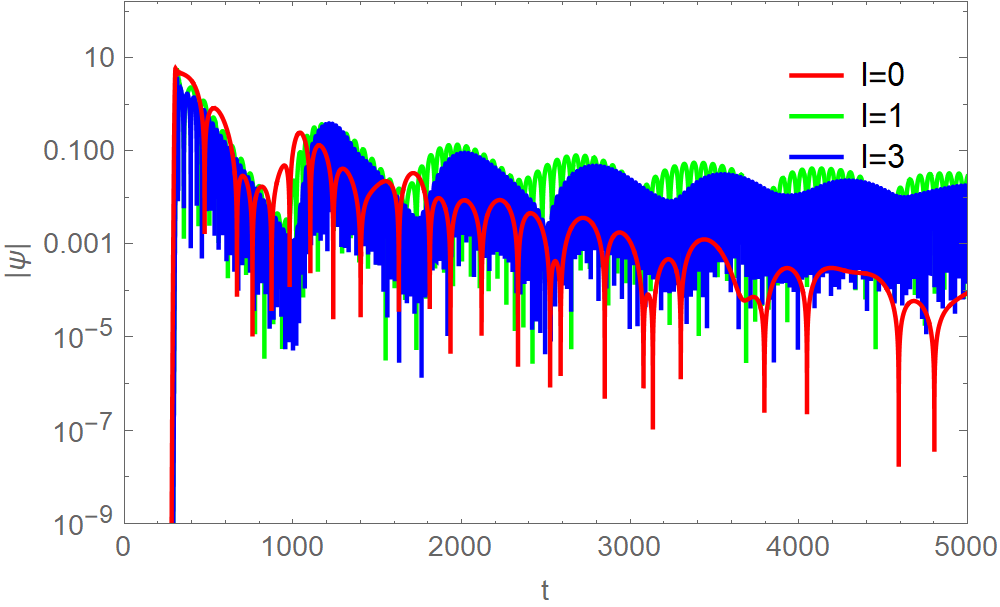}
}
\caption{\it\small Here are the time-domain profiles of varying $l$, associated with the double-peak potentials given in Fig.~\ref{changingl}. When $l=0$, the echo is hardly discernible, perhaps due to having a negative potential region.}\label{lecho}
\end{figure}
\begin{figure}[htbp]
\centering
\subfigure[$h(r)$]{
\includegraphics[width=0.45\textwidth]{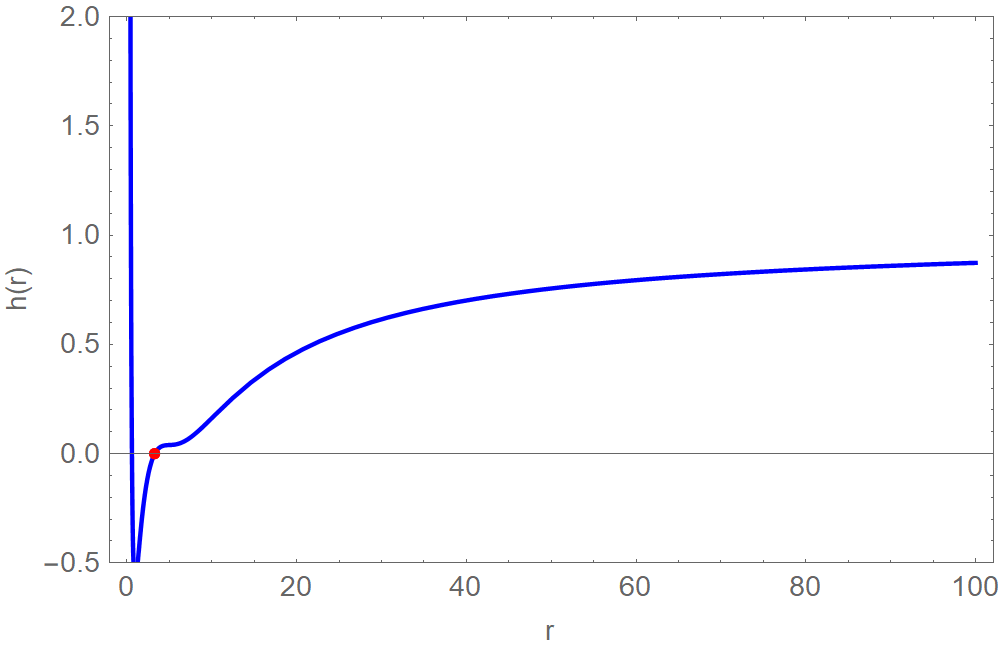}
}
\quad
\subfigure[$h'(r)$]{
\includegraphics[width=0.45\textwidth]{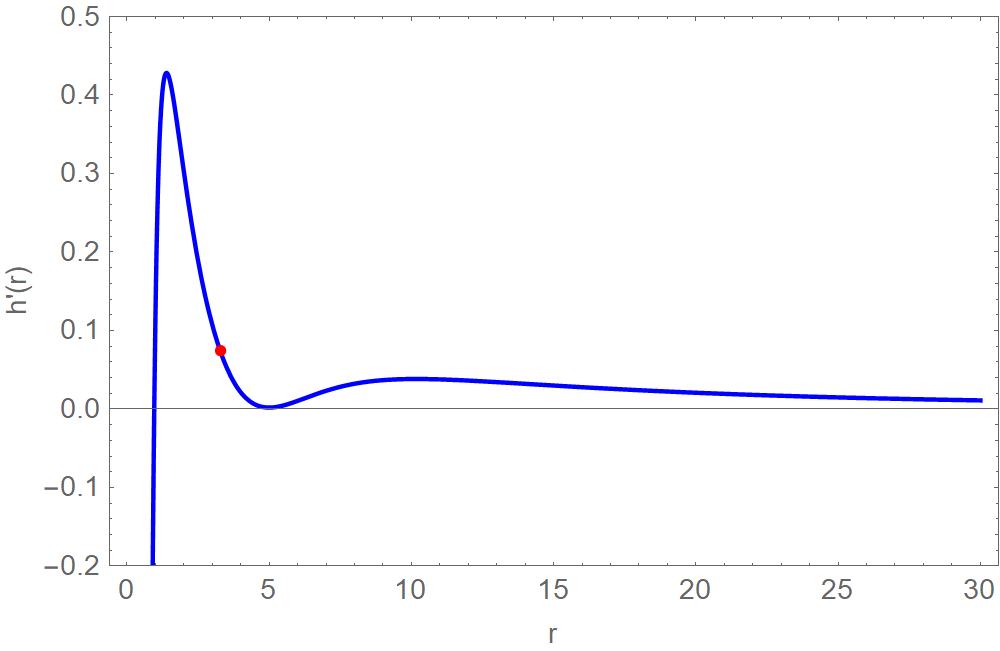}
}
\quad
\subfigure[$V(r_*)$]{
\includegraphics[width=0.45\textwidth]{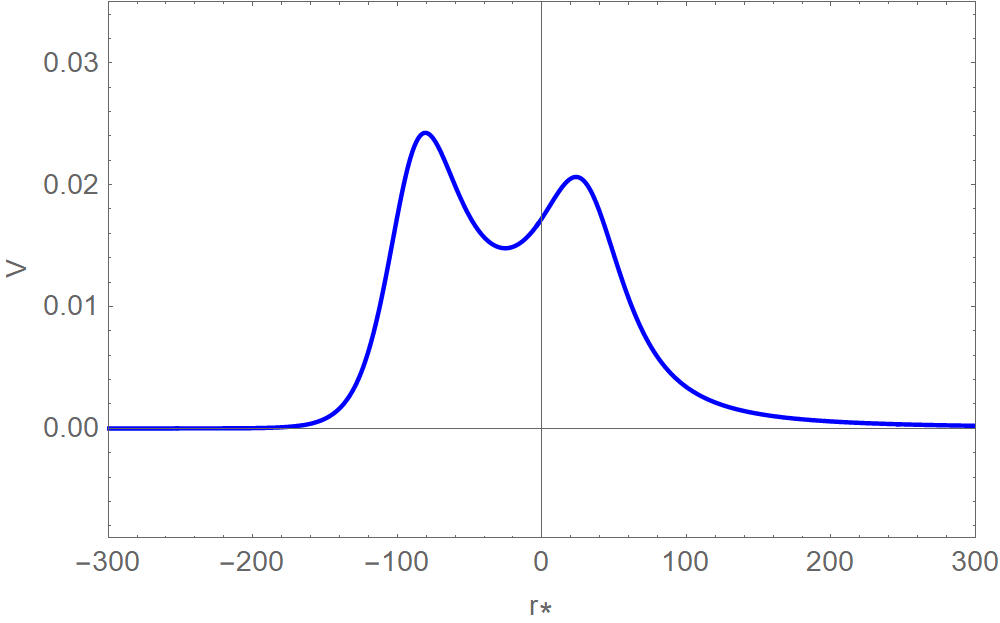}
}
\quad
\subfigure[Echoes]{
\includegraphics[width=0.45\textwidth]{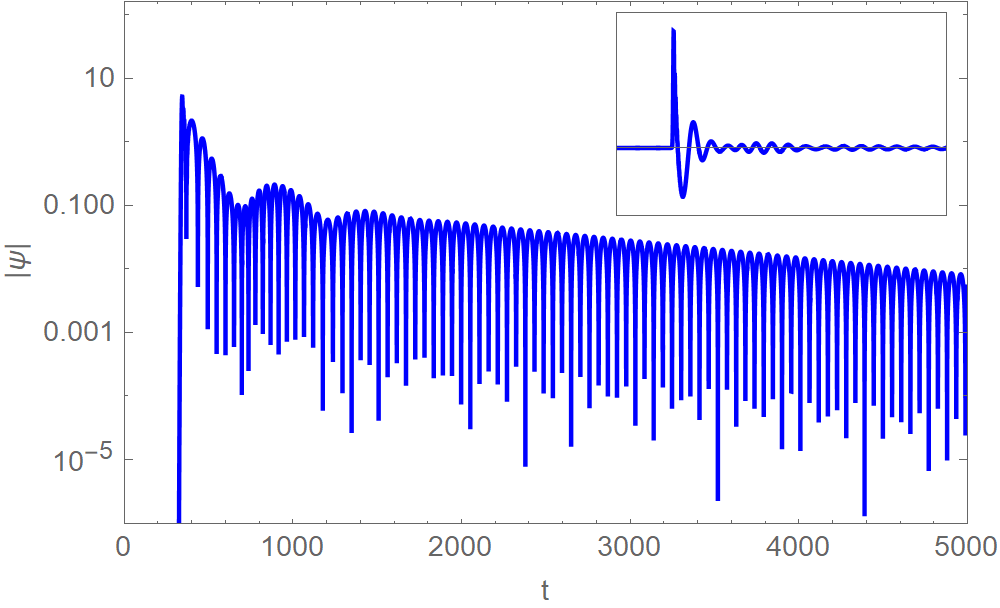}
}
\caption{\it\small Here we show that the double-peak potential can also arise even when the function $h$ is monotonous outside the horizon, provided that $h'$ is not. The black hole continue to satisfy DEC, but not SEC. The red point indicates the location the outer horizon. This black hole can also produce echoes, although they become indiscernible in later time. Here we set $\alpha=123.90, p=1.181, q=6.863, l=3, M=6.58$.}\label{hdandiao}
\end{figure}

\end{document}